\documentclass[preprint,showpacs,amsmath,amssymb]{revtex4}
\usepackage{epsfig}
\usepackage{rotating}
\usepackage{array}
\usepackage{epsfig}
\usepackage{graphicx}
\usepackage{epstopdf}
\usepackage{amssymb}
\usepackage{latexsym}

\newcommand{\be}{\begin{eqnarray}}
\newcommand{\bb}{\bibitem}
\newcommand{\ee}{\end{eqnarray}}

\newcommand{\fig}{\begin{figure}}
\newcommand{\ef}{\end{figure}}
\newcommand{\bc}{\begin{center}}
\newcommand{\ec}{\end{center}}

\newcommand{\bn}{\begin{enumerate}}
\newcommand{\en}{\end{enumerate}}

\newcommand{\bz}{\begin{itemize}}
\newcommand{\ez}{\end{itemize}}
\newcommand{\ct}{\centerline}
\newcommand{\ep}{\epsfig}
\newcommand{\cp}{\caption}

\newcommand{\ba}{\begin{array}} 
\newcommand{\ea}{\end{array}}

\newcommand{\bt}{\begin{tabular}}
\newcommand{\et}{\end{tabular}}

\newcommand{\mc}{\mathcal}

\newcommand{\bd}{\begin{displaymath}}
\newcommand{\ed}{\end{displaymath}}
\newcommand{\nn}{\nonumber}
\newcommand{\ben}{\begin{eqnarray*}}
\newcommand{\een}{\end{eqnarray*}}

\newcommand{\al}{\alpha}
\newcommand{\bet}{\beta}

\newcommand{\Ga}{\Gamma}

\newcommand{\bq}{\begin{quote}}
\newcommand{\eq}{\end{quote}}

\begin{document}
\def\bea{\begin{eqnarray}}
\def\eea{\end{eqnarray}}
\def\nn{\nonumber}
\newcommand{\snu}{\tilde \nu}
\newcommand{\sll}{\tilde{l}}
\newcommand{\asnu}{\bar{\tilde \nu}}
\newcommand{\stau}{\tilde \tau}
\newcommand{\dmsnu}{{\mbox{$\Delta m_{\tilde \nu}$}}}
\newcommand{\mt}{{\mbox{$\tilde m$}}}

\renewcommand\epsilon{\varepsilon}
\def\be{\begin{eqnarray}}
\def\ee{\end{eqnarray}}
\def\lla{\left\langle}
\def\rra{\right\rangle}
\def\za{\alpha}
\def\zb{\beta}
\def\lsim{\mathrel{\raise.3ex\hbox{$<$\kern-.75em\lower1ex\hbox{$\sim$}}} }
\def\gsim{\mathrel{\raise.3ex\hbox{$>$\kern-.75em\lower1ex\hbox{$\sim$}}} }
\newcommand{\Rbs}{\mbox{${{\scriptstyle \not}{\scriptscriptstyle R}}$}}

\draft


\title{Doubly Coexisting Dark Matter Candidates\\
in an Extended Seesaw Model}


\thispagestyle{empty}
\author{  Sin Kyu Kang$^{1,}$\footnote{E-mail:
        skkang@snut.ac.kr},~~ H. Sung Cheon$^{2,}$\footnote{E-mail:
        hscheon@gmail.com},~~ C. S. Kim$^{2,}$\footnote{E-mail:
        cskim@yonsei.ac.kr} }
\affiliation{ $ ^{1}$ School of Liberal Arts, Seoul National University of Technology,
                     Seoul 121-742, Korea \\
              $ ^{2}$ Dept. of Physics and IPAP, Yonsei University, Seoul 120-749, Korea }

\begin{abstract}
\noindent We examine how a scenario of coexisting two-particle dark mater can be realized
in the extended seesaw model,
which we have proposed previously to accommodate small neutrino masses and low scale leptogenesis
with an introduction of singlet Majorana neutrino $S$ and singlet scalar $\phi$.
We now impose the discrete symmetry $Z_2 \times Z_2^{\prime}$
and introduce new renormalizable interaction terms with a new heavy singlet scalar particle $\Phi$
so as for previously introduced $S$ and  $\phi$ to be doubly coexisting dark matter candidates.
Depending on the mass spectrum of the two dark matter candidates, the annihilation process either
$SS\longrightarrow \phi \phi$ or $\phi\phi\longrightarrow SS$ is of particular interest because
the annihilation cross sections for the processes can be so large that the relic abundance of
decaying particle should get lowered,
which in turn makes the constraints on its parameter space relaxed,
compared with the case of one and only one dark matter candidate.
We discuss the implications of the dark matter detection through the scattering
off the nucleus of the detecting material on our scenarios for dark matter candidates.
We also study the implications for the search
of invisible Higgs decay at LHC, which may serve as a probe of our scenario for dark matter.

\end{abstract}
\maketitle \thispagestyle{empty}

\section{Introduction}

To identify the nature of dark matter remains an open question in particle physics and cosmology.
The amount of cold dark matter in the Universe, which has been determined precisely
from 5 year WMAP data \cite{WMAP}, is given by $\Omega_{\rm CDM}h^2 = 0.1099 \pm 0.0062$.
None of the SM particles can be a good candidate for the dark matter, so that the existence of
the dark matter itself points to new physics beyond the Standard Model (SM).
It has been argued that weakly interacting massive particles (WIMPs) are promising dark matter candidates among various proposals in the literature \cite{DM1, DM2}.
With the appropriate size of its interactions, natural values
of its mass around the TeV scale and annihilation cross-section into the SM particles,
the observed cosmological abundance is suitably accounted for by the WIMP relic density \cite{DM2, WIMP}.
This weak-scale annihilation cross-section, when reversed, properly suggests a weak-scale production cross-section at colliders,
and, when viewed in the $t$-channel, implies an elastic scattering cross-section on nuclei
is within reach of purpose-built underground detectors. In recent years, experiments
of the latter type have reached an impressive level of sensitivity \cite{xenon100, scdms}, and significant future
progress in this direction is anticipated.

While dark matter is usually assumed to have one and only one particle candidate,
a possibility of multiply coexisting dark matter particles has been recently explored \cite{ma}
(There has been an attempt to demonstrate the total amount of dark matter 
with a composition of keV sterile neutrinos and a generic CDM \cite{sterile}.)
The simplest generic scenario of the multi-particle dark matter is to have coexisting two particles
as dark matter candidates by imposing discrete symmetry $Z_2 \times Z^{\prime}_2$ in the Lagrangian.
As discussed in Ref. \cite{ma}, the scenario can be naturally realized in the minimal supersymmetric
Standard Model.
One of virtues of such a scenario is that severe constraints on the model parameter space can be
relaxed due to the presence of the second dark matter particle.

Recently, we proposed a model to accommodate low scale leptogenesis, tiny neutrino masses
and the existence of dark matter \cite{ckk}. Our aim has been accomplished by introducing new gauge singlet neutrinos and
a singlet scalar boson on top of the SM particles and the right handed singlet Majorana neutrinos
\cite{kk}.
Imposing discrete $Z_2$ symmetry in the model, either the newly introduced singlet neutrino or
the singlet scalar boson could be a dark matter candidate depending on their mass spectrum under
the assumption that dark matter is one and only one particle candidate as usual \cite{singletDM, self, sh}.
Motivated by the work \cite{ma}, in this paper, we examine if the new singlet neutrino and the singlet scalar boson in the model
can be coexisting two-particle dark matter (2DM) candidates abandoning the assumption on only one dark matter.
In fact, as discussed in \cite{ckk}, when the new singlet neutrino in the model we proposed is the only
one  dark matter candidate,
the co-annihilation process \cite{coan} is responsible for the relic abundance required for dark matter and
the allowed parameter space for achieving the right amount of the relic abundance has been
determined to be very narrow.
Thus, it would be interesting to investigate whether the 2DM scenario realized in our model can make the allowed parameter space relaxed or not.
In this scenario, we introduce new renoramlizable interaction terms with a new heavy singlet scalar
particle $\Phi$ so as for previously introduced $S$ and $\phi$ to be doubly coexisting dark matter candidates.
As will be discussed later, the important point deserved to notice is
that such scenario may open up a new annihilation process of the heavier dark matters
into the lighter ones, which dominantly contributes to the relic abundance and thus
plays a crucial role in making the allowed parameter space relaxed.

This paper is organized as follows:
In Section II, we show how the scenario of coexisting 2DM candidates can be realized
in the model we previously proposed. To achieve our goal, we introduce new renormalizable
interaction terms among
the SM Higgs boson, the heavy singlet scalar, the singlet neutrino and the singlet scalar boson
in the Lagrangian.
In Section III, we investigate the relic abundance and present the allowed regions
of the parameter space for the possible coexisting 2DM candidates.
In Section IV, we discuss about dark matter detection through the scattering off the nucleus of the detecting material.
In Section V, we study how we can probe the coexisting 2DM candidates through the search for the
 Higgs decay at collider experiment, particularly at LHC.

\section{coexisting two-particle dark matter candidates}

To accommodate low scale leptogenesis, tiny neutrino masses and dark matter, the model we proposed in the previous work \cite{ckk}
is described by the following Lagrangian,
\be
\mc{L}_{{\rm Ref.}~[9]} &=& \mc{L}_{\rm SM} +(Y_D \bar{\nu} H N + Y_S \bar{N} \phi S +h.c.)+ M_R N^{T}N - m_{S^0} S^{T}S \nn \\
&& +\frac{1}{2}(\partial_\mu \phi)^2 - \frac{1}{2}m^2 _{\phi^0}\phi^2 -
\frac{\lambda_s}{4} \phi^4 - \lambda H^\dagger H \phi^2 ,\label{lag1}  \ee
where the first term is the Lagrangian of the SM and $\nu$, $N$, $S$ stand for $SU(2)_L$ doublet,
right-handed singlet and singlet Majorana neutrinos, respectively. $H$ and $\phi$ denote the $SU(2)_L$ doublet and singlet scalar fields.
As discussed in \cite{ckk}, the newly introduced neutral particle, either singlet Majorana neutrino or singlet scalar
boson, can be a candidate for dark matter, provided that one imposes $Z_2$ symmetry under which  $S$ and $\phi$ are odd and all other particles even.
In this model, low scale leptogenesis of order  1-10 TeV can be achieved when we take, for example, $Y_D \sim 10^{-6}$ and $Y_S \sim 10^{-3}$ \cite{kk}.
Due to such small Yukawa couplings , the cross section for the annihilation of $S$ into a pair of $\phi$ is too small,
so that the coannihilation processes are compulsory to achieve the right amount of the relic abundance, which in turn lead to
a bit tight constraints on the parameter space \cite{ckk}.

Now, let us extend the model described by the Lagrangian Eq. (\ref{lag1}) so that the scenario of
the coexisting 2DM can be realized. In order to guarantee the stability
of the 2DM candidates, we first impose the discrete symmetry $Z_2 \times Z_2^{\prime}$
under which all the SM particles are $(+,+)$,
singlet neutrino $S$ is $\sim (-,+)$ and the singlet scalar boson $\phi$ is $\sim(+,-)$.
In addition to the Lagrangian Eq. (\ref{lag1}), we introduce new renormalizable  terms given by
\be
\mc{L} = &&\mc{L}_{{\rm Ref.}~[9]} + Y_\Phi \bar{S} \Phi S +\frac{1}{2} m^2 _{H^0} H^\dagger H - \frac{\lambda_1}{4} H^\dagger H H^\dagger H + \frac{1}{2}m^2 _{\Phi^0} \Phi^2\nn \\
&-& \frac{\lambda_2}{4} \Phi^4 - \lambda_3 \phi^2 \Phi^2 - \lambda_4 H^\dagger H \Phi^2, \label{lag2}
\ee
where $\Phi$ is the SM-like (+,+) heavy singlet scalar particle, whose mass is assumed to be
larger than those of $\phi$ and $S$.
Here, we demand that the minimum of the scalar potential is bounded from below so
as to guarantee the existence of vacuum
and the minimum of the scalar potential must spontaneously
break the electroweak gauge group, $<H^0>, <\Phi> \neq 0$, but must not break
$Z_2 \times Z_2 '$ symmetry imposed above.
%
%
After spontaneous symmetry breaking, the part of the scalar potential is given by
\be V=&&\frac{1}{2} m^2 _\phi \phi^2 - \frac{1}{2} m^2 _{h} h^2 -\frac{1}{2}m^2 _\Phi \Phi^2 + 2 \lambda_4 v_h v_\Phi h\Phi + \frac{\lambda_s}{4} \phi^4 +\frac{\lambda_1}{4}v_h h^3+\frac{\lambda_1}{16} h^4  \nn \\
&+& \frac{\lambda_2}{4} \Phi^4 + \lambda_2 v_\Phi \Phi^3 + \frac{\lambda}{2}\phi^2 h^2 + \lambda v_h\phi^2 + \lambda_3 \phi^2 \Phi^2 +2 \lambda_3 v_\Phi \Phi \phi^2 \nn \\
&+&\frac{\lambda_4}{2}h^2\Phi^2+\lambda_4 v_\Phi h^2 \Phi + \lambda_4 v_h\Phi^2 + h.c., \label{pot}
\ee
where
\be m^2 _\phi &=& m^2_{\phi^0} +\lambda v^2 _h + 2\lambda_3 v^2 _{\Phi} \nn \\
m^2 _h &=& \frac{1}{2}m^2 _{H^0} -\frac{3}{4}
\lambda_1v^2 _h -\lambda_4 v_\Phi ^2 \nn \\
m^2 _\Phi &=& m^2 _{\Phi^0} - 3 \lambda_2 v^2 _\Phi - \lambda_4 v^2 _h  \label{pmass}
\ee
Here, we have adopted $\sqrt{2}H^T = (h,0)$ and shifted the Higgs boson $h$
and the singlet scalar  $\Phi$ by $h\rightarrow h+v_h $ and $\Phi \rightarrow \Phi + v_\Phi$, respectively.
The relevant size of $v_\Phi$ in this work
is of order $1$ TeV.
Since there exists a mixing mass term between $h$ and $\Phi$,
we rotate them with $\Phi =-sh'+ c\Phi'$ and $h=ch'+s\Phi'$,
where $s$ and $c$ are  $\sin\theta$ and $\cos\theta$, respectively.
Then, the effective potential is written as
\be V_{eff} =&& \frac{m_\phi '^2}{2}\phi^2 + \frac{m_h '^2}{2}h'^2 + \frac{m_\Phi '^2}{2}\Phi'^2 +\frac{\lambda_4}{2}h'^2 \Phi'^2 \nn \\
&+& \frac{1}{2}(\lambda c^2+2\lambda_3 s^2) \phi^2 h'^2+(\frac{\lambda}{2}s^2 +\lambda_3 c^2)\phi^2 \Phi'^2 \nn \\
&+& \kappa_1 h'^3  + \kappa_2 \Phi'^3  + \al \Phi' \phi^2+\bet h'^2 \Phi' + \gamma h' \phi^2 \nn \\
&+& \Big (\frac{\lambda_1}{16}c^4+\frac{\lambda_4}{2}c^2s^2+\frac{\lambda_2}{4}s^4 \Big  ) h'^4
  + \Big (\frac{\lambda_2}{4 }c^4+\frac{\lambda_4}{2}c^2s^2+\frac{\lambda_1}{16}s^4 \Big ) \Phi'^4  , \label{pot2}
\ee
where
\be
\kappa_1 &=& \lambda_4 c s (cv_\Phi + s v_h)+\frac{\lambda_1}{4}c^3v_h  + \lambda_2 v_\Phi s^3 \nn \\
\kappa_2 &=& \lambda_2 c^3 v_\Phi -\lambda_4 cs(cv_h  - sv_\Phi)-\frac{\lambda_1}{4}s^3 v_h  \nn \\
\al &=& -\lambda v_h  s +2\lambda_3 v_\Phi c \nn \\
\bet &=&  \lambda_4 [ v_\Phi c^3 - v_h s^3 + 2sc (s v_\Phi- cv_h )]
+cs[-\frac{3}{4}\lambda_1 v_h  c +3\lambda_2 v_\Phi s ]\nn \\
\gamma &=& 2\lambda_3 s v_{\Phi} + \lambda v_h c .
\ee
On the other hand, the Lagrangian containing the field $S$ becomes,
\be \mc{L}_S = - m_S S^T S +(s Y_\Phi)  h' S^T S + (c Y_\Phi)  \Phi' S^T S, \label{ls2}
\ee
where $m_S=(m_{S^0} - Y_\Phi v_\Phi)$ is the physical mass of S.

For our purpose, in this work, we keep the new coupling constants $Y_\Phi$, $\lambda_{i (i=1,..,4)}$
to be non-zero.
For $m_S \gtrsim m_{\phi}$, we see from the interaction terms in Eqs. (\ref{pot2}, \ref{ls2})
that  the singlet neutrino $S$
can annihilate into $\phi \phi$ and  $h h$, in addition to the ordinary particles.
The produced scalar field $\phi$ from the annihilation of $S$ decays into the SM particles.
The annihilation process, $S S\rightarrow \phi  \phi$, is of particular interest because
the annihilation cross section for this process can be so large that $S$ could predominantly
annihilate into $\phi\phi$, resulting in a much
smaller relic abundance, thereby relaxing the constraints on its parameter space as expected.
In fact, the annihilation of $S$ into $\phi \phi$ also occurs in the scenario
with only one dark matter candidate proposed in \cite{ckk}, but its contribution to the relic abundance
can not work because both $S$ and $\phi$ are odd particles under $Z_2$ symmetry.
Thus, the enhancement of the annihilation rate for the process $S S\rightarrow \phi  \phi$
is a distinctive feature of the 2DM scenario.
For $m_{\phi} \gtrsim m_S$, the conclusions are mostly the same if we switch $S$ and $\phi$.

\section{Relic abundance in two-particle dark matter scenario}

Assuming the coexistence of two dark matter candidates, the relic abundance observed must
be composed of the contributions of both $S$ and $\phi$ as follows,
\be \Omega_S h^2 + \Omega_\phi h^2 = \Omega_{\rm CDM} h^2 = 0.110 \pm 0.006.
\ee
The relic density of each dark matter species is approximately given by
$$\Omega_i h^2\approx (0.1 pb)/ <\sigma v>_i~~~~~~~(i=S,\phi),$$
where $<\sigma v>_i$ is the thermally averaged product of its annihilation cross section with its velocity.
For our convenience, we define the parameter $\epsilon_i$ as a ratio of $\Omega_i h^2$ to
$\Omega_{\rm CDM} h^2$,
\be \epsilon_i = \frac{\Omega_i h^2}{\Omega_{\rm CDM} h^2}~, \ee
where $\epsilon_S+\epsilon_{\phi}=1$.
In fact, the parameter $\epsilon_i$ represents the fraction of the mass density of each dark matter
species in our local dark-matter halo as well as in the Universe.
Since the values of $\epsilon_i$ are unknown, we consider a few cases by choosing their values
in the analysis.

To calculate $\Omega_i h^2$, the input parameters to be fixed are
the SM Higgs mass $m_h$ and masses of $m_S$, $m_{\phi}$ and $m_{\Phi}$,
the vacuum expectation values $v_h $ and $v_\Phi$, mixing angle $\theta$ and
the coupling constants $Y_\Phi$, $\lambda$, $\lambda_{i(i=1,...,4)}$.
First of all, varying the two parameters $Y_\Phi$ and $m_S$ while fixing all others,
we calculate each $\Omega_i h^2$ with the help of the \emph{micrOMEGA}s 2.0.7 program \cite{micro},
and then pick up the parameter space $Y_\Phi$ and $m_S$ resulting in the chosen value
of $\epsilon_S $.

\fig [t] \ct{\ep{figure=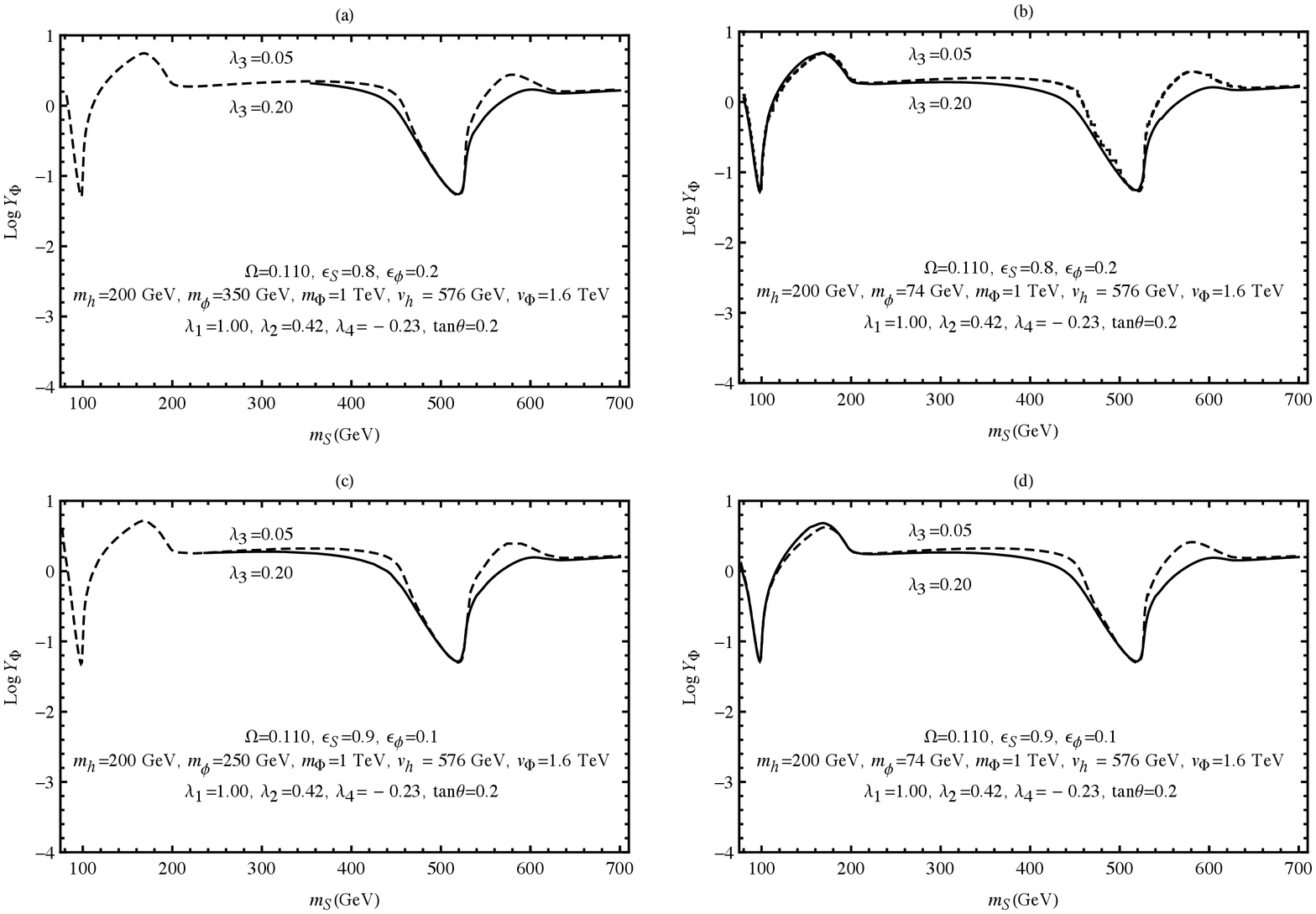, scale=0.57}}
\cp{Plots of $Y_\Phi$ as a function of $m_S$
for $\Omega_{\rm CDM} h^2 = 0.110$.
We take $m_h=200$ GeV.
The panels correspond to (a)
$m_\phi = 350$ GeV, $\epsilon_S = 0.8,\, \epsilon_\phi = 0.2$,
(b) $m_\phi = 74$ GeV, $\epsilon_S = 0.8,\, \epsilon_\phi = 0.2$,
(c) $m_\phi = 250$ GeV, $\epsilon_S = 0.9,\, \epsilon_\phi = 0.1$,
(d) $m_\phi = 74$ GeV, $\epsilon_S = 0.9,\, \epsilon_\phi = 0.1$.
The dashed (solid) lines correspond to $\lambda_3 = 0.05~ (0.20)$.}
\ef

In Fig. 1, we present the correlations between $Y_\Phi$ and $m_S$ for (a)  $m_{\phi}=
350$ GeV and $\epsilon_s=0.8$, (b) $m_{\phi}=
74$ GeV and $\epsilon_s=0.8$, (c)  $m_{\phi}=
250$ GeV and $\epsilon_s=0.9$, (d)  $m_{\phi}=
74$ GeV and $\epsilon_s=0.9$.
The solid and dashed lines correspond to $\lambda_3=0.20$, $\lambda_3=0.05$, respectively.
Here, we take the value of $\Omega_{\rm CDM} h^2$ to be 0.110, and the input values of the other parameters
are given by  $m_h=200$ GeV, $m_{\Phi}=1$ TeV, $\lambda_1=1.0$, $\lambda_2 =  0.42$, $\lambda_4 = -0.23$, $v_h = 576$ GeV, $v_\Phi = 1.6 $ TeV  and $\tan\theta=0.2$.
For the regime $m_{\phi} \lesssim m_S$, the relic abundance of the field $S$ depends
on the annihilation of $S$ into the singlet scalar $\phi$ and the SM particles whose annihilation cross sections
contain the couplings  $\lambda$, $\lambda_{i(i=1,...,4)}$, $Y_\Phi$ as well as the SM Yukawa couplings.
In this case, the abundance of the singlet scalar field $\phi$ depends mainly on the annihilation
processes  of $\phi$ into the SM particles whose annihilation cross section
 contains the coupling $\lambda$, $\lambda_3$ and the SM Yukawa couplings.
 Thus, the value of $\lambda$ is determined from the chosen values of $\epsilon_{\phi}$ and $\lambda_3$.
 Once $\lambda$ is fixed in this way, we can obtain the correlation between
 $Y_\Phi$ and $m_S$ for a fixed value of $\lambda_3$ as presented in Fig. 1
by fitting to the chosen value of $\epsilon_S$.
On the other hand, for the regime $m_S \lesssim m_{\phi}$, the relic abundance of $S$ depends mainly
on the annihilation process of $S$ into the SM particles  whose annihilation cross sections
contain the coupling $Y_\Phi$ and the SM Yukawa couplings.
Thus, the value of $Y_\Phi$  {\bf as a function of $m_S$} is determined from the chosen value of $\epsilon_S$
irrespective of the value of $\lambda_3$.
We see from  Fig. 1 that the region  $m_S \lesssim m_{\phi}$ is not
 allowed in the case of $\lambda_3=0.2$.
This is because the rate of the annihilation of $\phi$ into $SS$ becomes large for $m_S \lesssim m_{\phi}$
and large value of $\lambda_3$, which makes $\epsilon_{\phi}$ lower for achieving the required relic abundance.

\fig [t] \ct{\ep{figure=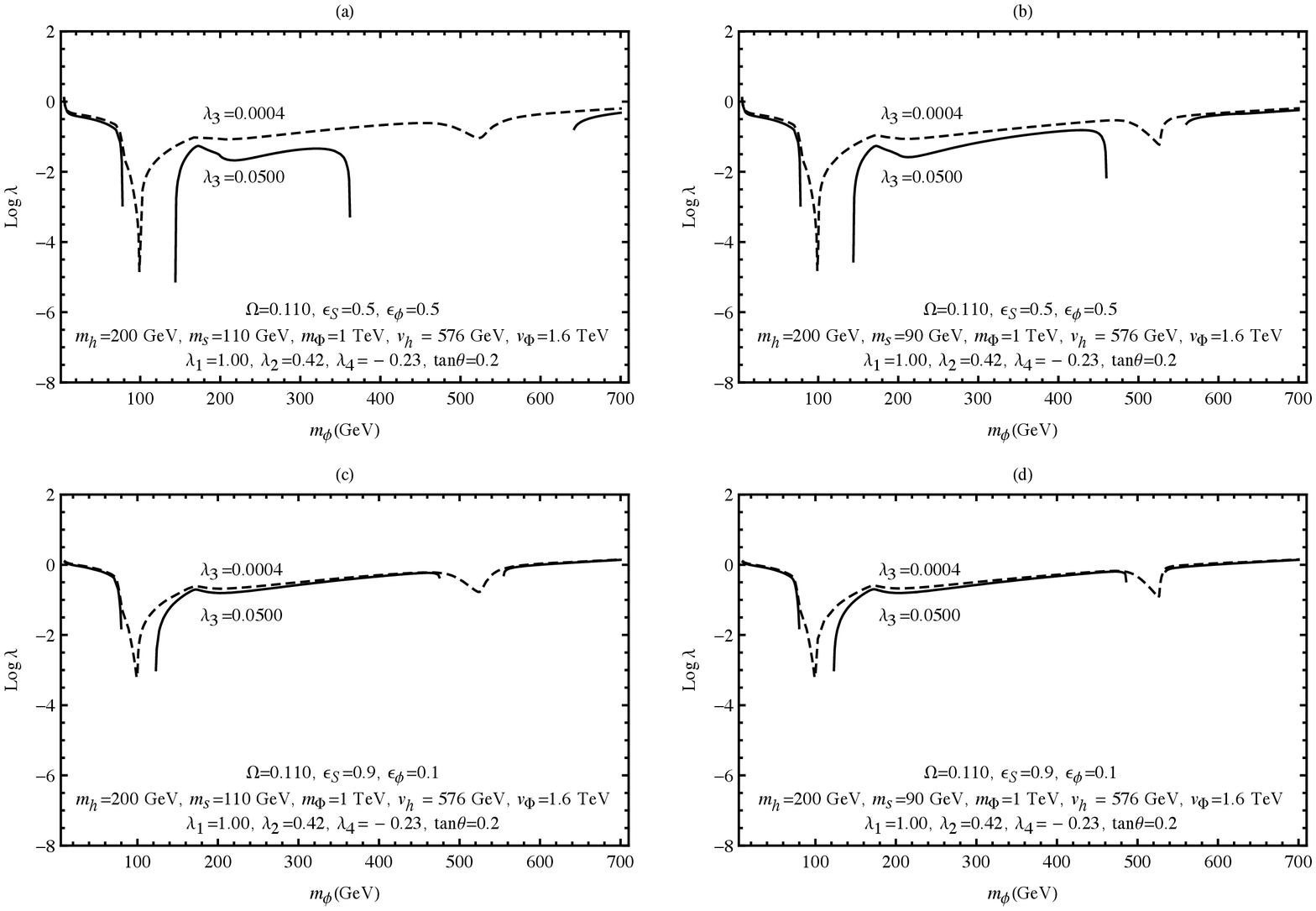, scale=0.6}}
\cp{Plots of $\lambda$ as a function of
$m_\phi$ for $\Omega_{\rm CDM} h^2 = 0.110$. We take $m_h = 200$ GeV
The panels correspond to (a) $m_S=110$ GeV, $\epsilon_S = 0.5$, $\epsilon_\phi = 0.5$,
(b) $m_S=90$ GeV, $\epsilon_S = 0.5$, $\epsilon_\phi = 0.5$,
(c) $m_S=110$ GeV, $\epsilon_S = 0.9$, $\epsilon_\phi = 0.1$ and (d) $m_S=90$ GeV,
$\epsilon_S = 0.9$, $\epsilon_\phi = 0.1$.
The dashed (solid) lines represent $\lambda_3 = 0.0004~(0.0500)$.} \ef

In Fig. 2, we present the correlation between $\lambda$ and $m_{\phi}$ for
$\Omega_{\rm CDM} h^2=0.110$ and (a) $m_S=110$ GeV and $\epsilon_{S}=0.5$, (b) $m_S=90$ GeV and $\epsilon_{S}=0.5$,
(c) $m_S=110$ GeV and $\epsilon_{S}=0.9$, and (d) $m_S=90$ GeV and $\epsilon_{S}=0.9$.
The solid (dashed) lines correspond to $\lambda_3=0.0500 ~(0.0004)$.
The values of the input parameters $m_h, m_{\Phi}, v_h, v_{\Phi}, \theta, \lambda_1, \lambda_2, \lambda_4$
are the same as those in Fig. 1.
For the regime $m_S \lesssim m_{\phi}$, the relic abundance of the field $\phi$ depends
on the annihilation of $\phi$ into the singlet fermion $S$ and the SM particles whose annihilation cross sections
contain the couplings $\lambda$, $\lambda_{i(i=1,...,4)}$, $Y_\Phi$ as well as the SM Yukawa couplings.
In this case, the abundance of the singlet neutrino $S$ depends on the annihilation
processes of $S$ into the SM particles whose annihilation cross section
 contains the coupling $Y_\Phi$ and the SM Yukawa couplings.
 Thus, the value of $Y_\Phi$ is first determined from a chosen value of $\epsilon_S$, and then
we  obtain the correlation between $\lambda$ and $m_{\phi}$ for a fixed value of $\lambda_3$,
as presented in Fig. 2
by fitting to the chosen value of $\epsilon_{\phi}$.
As one can see from Fig. 2, there exist two disconnected regions of $m_{\phi} $
in each panels for $\lambda_3 =0.05$: one is
around $2m_\phi\simeq m_h$ corresponding to the SM higgs resonance region,
and the other is around $2m_\phi \simeq m_\Phi$ corresponding to the $\Phi$ resonance region,
both of which are not allowed.
On the other hand, for the regime $m_{\phi} \lesssim m_S$, the relic abundance of $\phi$ depends
on the annihilation process of $\phi$ into the SM particles whose annihilation cross section
contains the coupling $\lambda$, $\lambda_3$ and the SM Yukawa couplings. Thus, the value of $\lambda$
as a function of $m_{\phi}$ is determined from the chosen values of $\epsilon_{\phi}$ and $\lambda_3$.

\section{Implication for dark matter search}

To directly detect dark matter, typically proposed method is to detect the scattering of
dark matter off the nucleus of the detecting material.
Since the scattering cross section is expected to be very small, the energy deposited
by a candidate for dark matter
on the detector nucleus is also very small. In order to measure this small recoil energy,
typically of order keV, of the nucleus,
a very low threshold detector condition is required.
Since the sensitivity of detectors to a dark matter candidate is controlled
by their elastic scattering cross section with nucleus,
it is instructive to examine how large the size of the elastic cross section could be.
First, to estimate the elastic cross section with nucleus, we need to know
the relevant matrix element for slowly moving spin-J nuclei,
which is approximately given \cite{sh} by
\be
\frac{1}{2J+1}\sum_{spins}|<n^{\prime}|\sum_{f} y_f \bar{f}f|n>|^2 \simeq \frac{|A_n|^2}{(2\pi)^6},
\ee
where $n$ denotes nucleons and $|A_n|$ is determined to be
\be
\mc{A}_{n} = g_{_{hnn}} \simeq \frac{190\, {\rm MeV}}{v_{_{\rm EW}}}
\label{an}
\ee
 by following the method given in \cite{sh} and taking the strange quark mass to be 95 MeV
 and $<n|\bar{s}s|n>\sim 0.7$.

\fig [b] \ct{\ep{figure=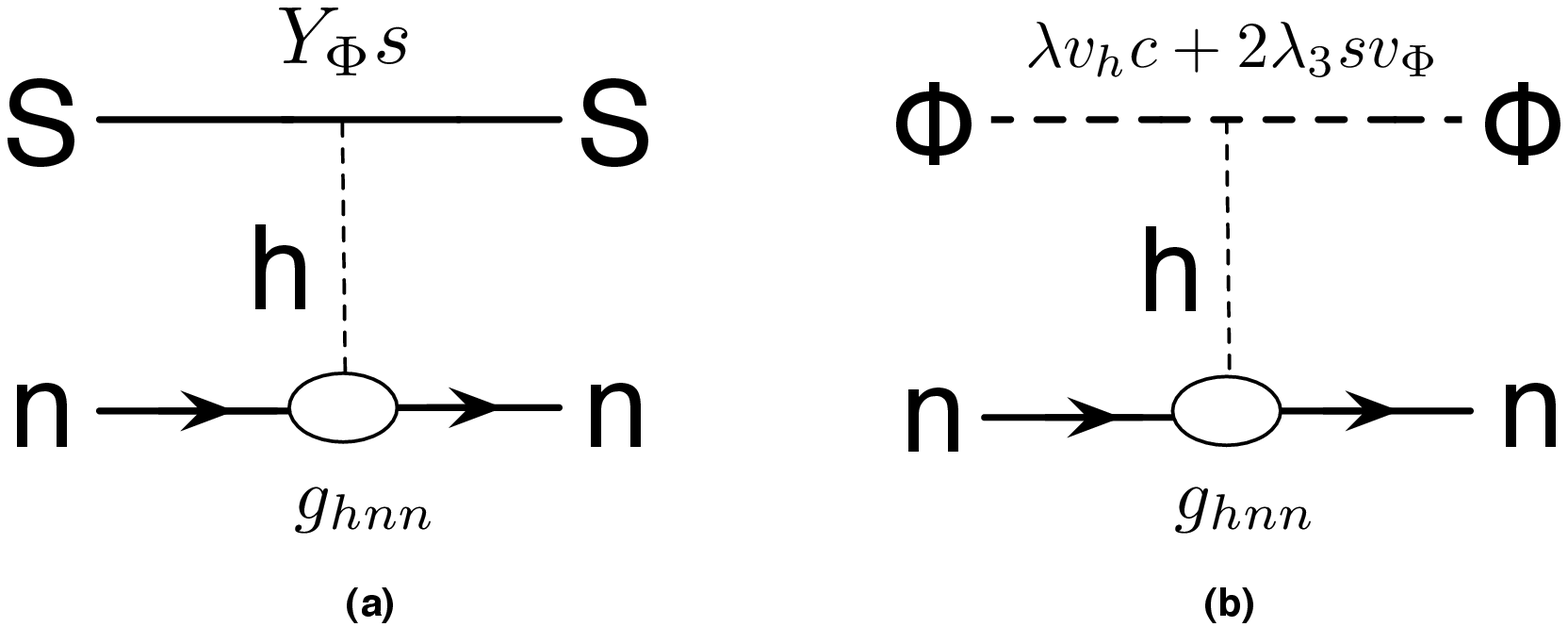, scale=0.65}} \cp{Feynman diagrams relevant to
(a) $S$-nucleon elastic scattering and
(b) $\phi$-nucleon elastic scattering.}
\ef

Now, let us estimate the sizes of the elastic scattering cross sections
in the scenario of the coexisting 2DM candidates, $S$ and $\phi$.
The Feynman diagrams relevant to dark matter-nucleon scattering are presented in Fig. 3.
The non-relativistic elastic scattering cross section for the singlet neutrino $S$-nucleon elastic
scattering is given by
\be
\sigma_{S} \approx \frac{(Y_\Phi s)^2|\mc{A}_n|^2}{\pi} \Big (\frac{m^2_*}{m^4 _h} \Big ), \label{el1}
\ee
where $m_* = m_S m_n / (m_S +m_n)$ is the reduced mass for the collision. Substituting Eq. (\ref{an})
into Eq. (\ref{el1}),
$\sigma_{S}({\rm nucleon})$ becomes
\be \sigma_{S}({\rm nucleon}) \approx \frac{1}{\pi}\Big (\frac{Y_\Phi s \times 190 \,
\mbox{MeV}}{m^2 _h v_{_{\rm EW}}} \Big )^2
\Big ( \frac{m_p m_S}{m_p + m_S} \Big )^2,
\ee
where the mass of $m_p$ is a mass of proton.

\fig [tb]
\ct{\ep{figure=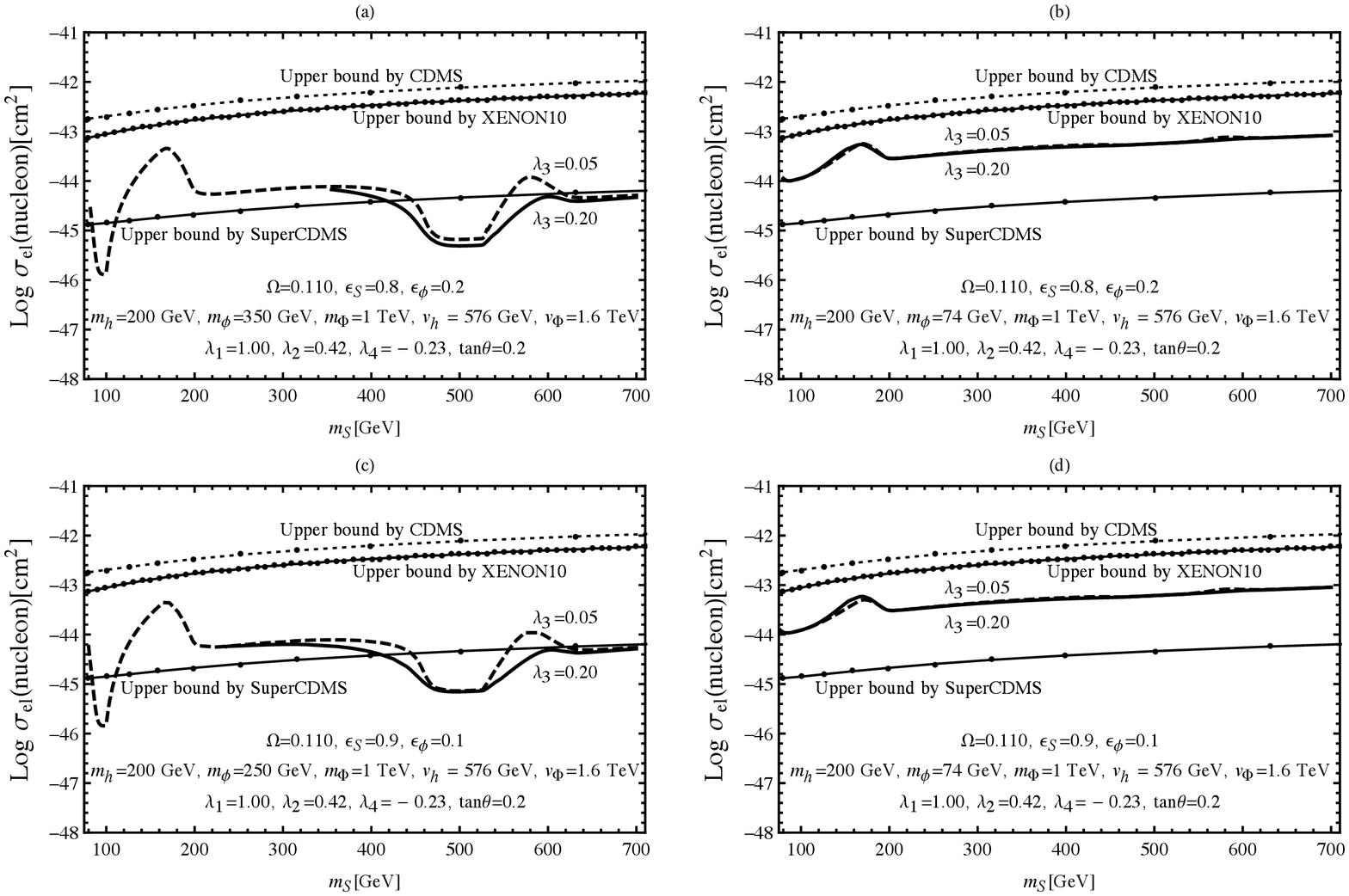, scale=0.6}}\cp{Plots of the elastic
cross section $\sigma_{el}$ as a function of $m_S$ for $m_h=200$ GeV with  $\epsilon_S =0.8$, $\epsilon_\phi =0.2$
and (a-b) $m_\phi=350~(74)$ GeV; with $\epsilon_S =0.9$, $\epsilon_\phi=0.1$
and (c-d) $m_\phi=250~(74)$ GeV.
We present that the spin-independent WIMP-nucleon cross section upper limits (90\% C.L.) by XENON10 Dark
Matter Experiment \cite{xenon} and CDMS experiment \cite{cdms}. In
addition the expected reach of SuperCDMS collaboration \cite{scdms}
is presented.} \ef

In the case of scalar $\phi$-nucleon elastic scattering,
the non-relativistic elastic scattering cross section for $\phi$ is given by
\be \sigma_{\phi} =
\frac{(\lambda v_h c + 2 \lambda_3 s v_\Phi)^2 |\mc{A}_n|^2}{4\pi}\Big ( \frac{m^2_*}{m^2 _\phi m^4 _h} \Big ),
\label{phielastic}
\ee
where $m_* = m_\phi m_n / (m_\phi + m_n)$ is the reduced mass for the collision.
Substituting (\ref{an}) into Eq. (\ref{phielastic}), $\sigma_{\phi}$(nucleon) becomes
\be
\sigma_{\phi} ({\rm nucleon}) \approx \frac{1}{4\pi}
\Big (\frac{(\lambda v_h c + 2\lambda_3 s v_\Phi) 190\, \mbox{MeV}}{m^2 _h v_{_{\rm EW}}} \Big )^2
\Big (\frac{m_p}{m_p + m_\phi} \Big )^2.
\label{phielastic2}
\ee

So far most experimental limits of the direct detection are given in terms of the scattering
cross section per nucleon under the assumption that there exists only one dark matter candidate.
In the scenario of 2DM, the cross section for the WIMP-nucleon elastic scattering $\sigma_{el}$ is
composed of the cross sections $\sigma_S$ and $\sigma_{\phi}$ \cite{ma};
\be \frac{\sigma_{el}}{m_0} = \frac{\epsilon_S}{m_S}\sigma_S + \frac{\epsilon_\phi}{m_\phi}\sigma_\phi,
\label{sigma}
\ee
where $m_0$ is the WIMP mass.
Without loss of generality, we  choose the singlet neutrino $S$ to be WIMP.
In Fig. 4, we plot the predictions for $\sigma_{el}$ as a function of $m_S$ for $m_h=200$ GeV in the cases of (a) $\epsilon_S =0.8$, $\epsilon_\phi =0.2$, $m_\phi=350$ GeV,
(b) $\epsilon_S =0.8$, $\epsilon_\phi =0.2$, $m_\phi=74$ GeV,
(c) $\epsilon_S =0.9$, $\epsilon_\phi =0.1$, $m_\phi=250$ GeV,
(d) $\epsilon_S =0.9$, $\epsilon_\phi =0.1$, $m_\phi=74$ GeV.

On calculating $\sigma_{el}$, we have used the correlation between $Y_\Phi$ and $m_S$ obtained in Fig. 1 corresponding to $\Omega=0.110$.
Here, we also plot the curves for the new $90\,\%$ C.L.
upper bounds on the WIMP-nucleon spin-independent cross
section as a function of WIMP mass obtained from XENON10 Dark Matter Experiment \cite{xenon}
and CDMS experiment \cite{cdms}.
In addition we present the expected reach of SuperCDMS collaboration \cite{scdms}.
While the current upper bounds on the spin-independent WIMP-nucleon cross sections
by XENON10 and CDMS Dark Matter Experiments
are not so strong to constrain the parameter space of the model,
the expected reach of SuperCDMS experiment would at least strongly constrain it.

\section{Implication for Higgs searches in collider experiments}

Now we investigate the implications of our new
scenario for Higgs searches in collider experiments. The singlet
scalar boson $\phi$ and singlet fermion $S$ will not directly couple to ordinary matters, but only
to the SM Higgs fields among the SM particles.
Therefore, although the presence of the singlet particles will not affect electroweak phenomenology
in a significant way, they
will affect the phenomenology of the Higgs boson. In the scenario of
only one dark matter candidate proposed in \cite{ckk}, the only
possible channel to probe the scenario through the Higgs search is
the invisible decay of the Higgs boson into $\phi \phi$ . However, in
the scenario of 2DM, the decay mode $h\rightarrow S S$ is also allowed
due to new interaction terms \cite{Khlopov}. The real Higgs boson can decay into a
pair of singlet scalars $\phi$ if $2m_{\phi}<m_h$, whereas it can
decay into a pair of singlet neutrinos $S$ if $2m_{S}<m_h$ . The
invisible Higgs decay widths are given at tree level by
\be
\Ga_{h\rightarrow \phi\phi} &=& \frac{(\lambda v_h c + 2\lambda_3 s v_\Phi)^2}{32 \pi m_h}
 \Big (1- \frac{4m^2 _\phi}{m^2 _h} \Big )^{1/2} ~~~~~(\mbox{for}~~ 2m_{\phi} < m_h), \\
{\rm and}~~~~~\Ga_{h\rightarrow SS} &=& \frac{(Y_\Phi s)^2 m_h}{8\pi}
 \Big (1 - \frac{4m_S ^2}{m_h ^2} \Big )^{3/2} ~~~~~~~~~\quad\quad (\mbox{for}~~ 2m_S < m_h).
\ee
If the Higgs boson mass $m_h$ is smaller than $2 m_{\phi(S)}$, then
the singlet particles can not be produced by real Higgs decays, but arise only via virtual Higgs exchange.
We notice that since any produced singlet particles are not expected to interact inside the collider,
they only give rise to strong missing energy signals.

To quantify the signals for the invisible decay of the Higgs boson,
we investigate the ratio $R$ defined as follows \cite{sh}:
\be
R=\frac{{\cal B}_{h\rightarrow W^{+}W^{-},ZZ,\bar{b}b,\, \bar{c}c, \, \bar{\tau}\tau} ({\rm SM}+\phi, S)}
 {{\cal B}_{h\rightarrow W^{+}W^{-},ZZ, \bar{b}b,\, \bar{c}c, \, \bar{\tau}\tau} ({\rm SM})} =
\frac{\Ga_{h,total} ({\rm SM})}{\Ga_{h\rightarrow \phi \phi(SS)} + \Ga_{h, total} ({\rm SM})}~. \label{Rratio}
\ee
The ratio $R$ indicates how the expected signal for the visible decay of the Higgs boson can decrease due to the existence of the singlet scalar field $\phi$ and/or neutrino $S$.
For $2 m_{\phi} > m_h$, the decay
mode $h\rightarrow \phi \phi$ is forbidden, so the ratio $R$ defined
in Eq. (\ref{Rratio}) becomes
\be R &=&
\frac{\Ga_{h,total}({\rm SM})}{\Ga_{h,total}({\rm SM})} = 1~~~~~ \qquad\qquad\,\, (2m_S \geq m_h) \nn \\
&=& \frac{\Ga_{h,total}({\rm SM})}{\Ga_{h\rightarrow SS} +\Ga_{h,total}({\rm SM})} ~~~~~~\qquad (2m_S <m_h)~.
\ee
On the other hand, for $ 2 m_{\phi} < m_h $, the decay mode $h\rightarrow \phi \phi$ is allowed,
so the ratio $R$ is given by
\be
R &=& \frac{\Ga_{h,total}({\rm SM})}{\Ga_{h\rightarrow \phi \phi} +
\Ga_{h,total}({\rm SM})}~~~~~ \qquad\qquad\,\, (2m_S \geq m_h) \nn \\
&=& \frac{\Ga_{h,total}({\rm SM})}{\Ga_{h\rightarrow S S}+\Ga_{h\rightarrow \phi \phi} +
\Ga_{h,total}({\rm SM})} \qquad (2m_S <m_h)~.
\ee

\fig [tb]
\ct{\ep{figure=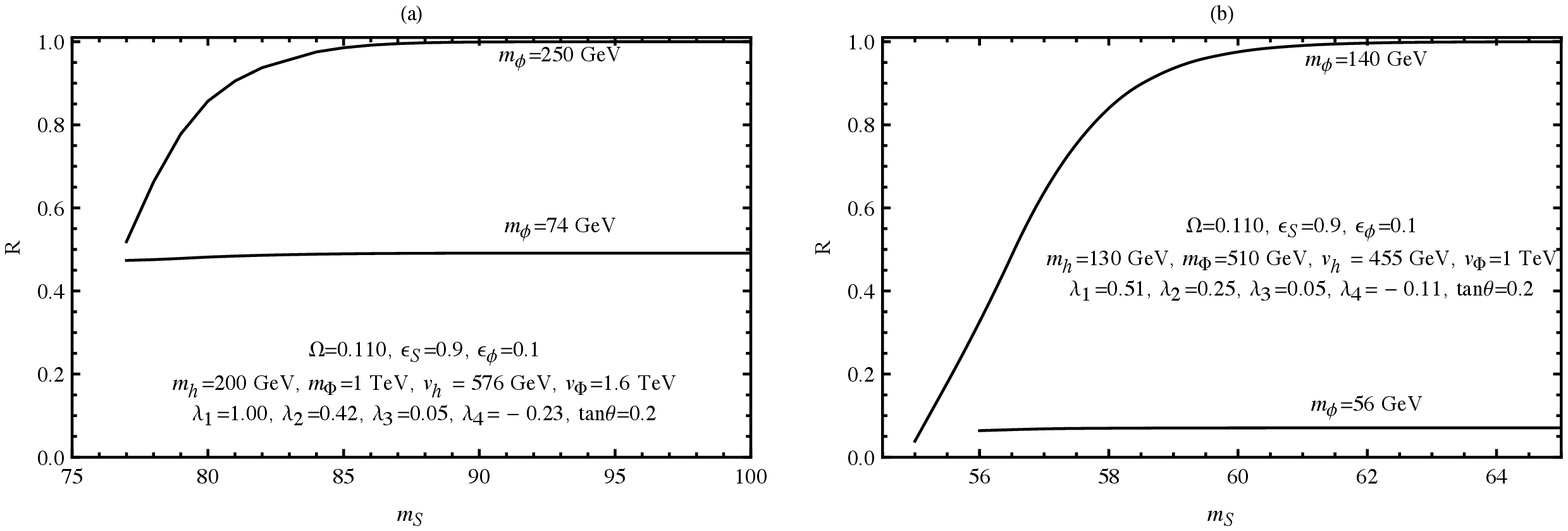, scale=0.6}}
\cp{Plots of the ratio R as a function of $m_S$ for our new model with
$\Omega h^2 = 0.110$ for (a)  $\epsilon_S =1$,
$\epsilon_\phi = 0$, $m_h = 200$ GeV and
(b)  $\epsilon_S=0.9$, $\epsilon_\phi=0.1$, $m_h = 130$ GeV.
The upper curves correspond to $m_{\phi}=$ (a) 120 GeV and (b) 90 GeV.
The lower curves correspond to $m_{\phi}=$ (a) 90 GeV and (b) 58 GeV.
}\ef

In Fig. 5,
we present the ratio $R$ as a function of $m_S$ for the cases  (a) $m_{\phi}=250~(74)$ GeV and
$m_h=200$ GeV,  (b) $m_{\phi}=140~(56)$ GeV and $m_h=130$ GeV,
which would be probed right after LHC starts to run.
The the upper and the lower curves in each panels of Fig. 5 correspond to $m_h < 2 m_{\phi}$ and $m_h > 2 m_{\phi}$,
respectively.
To estimate the ratio $R$, we need to specify some of unknown parameters:
$\Omega h^2 =0.110$ with $\lambda_3=0.05,~ \epsilon_S= 0.9, ~ \epsilon_{\phi}= 0.1$.
The upper limit of $m_S=100~(65)$ GeV, as shown in Fig. 5, is adopted from the condition $2m_S<m_h$.
Note that the upper curves  corresponding to a case for $2m_{\phi}>m_h$ indicate
that the scenario of 2DM can be probed via the ratio $R$ through the missing energy signal
only for $m_S < m_h/2$, $i.e.$ (a) $77~ \mbox{GeV} < m_S <100$ GeV and (b) $55~ \mbox{GeV} < m_S < 65~ \mbox{GeV}$.
The lower curves corresponding to a case for  $2m_{\phi}< m_h$ show that
the ratio $R$ is less than $\sim 0.5$ in (a) and $\sim 0.07$ in (b) for all possible values of $m_S$.
The reason why the value $R$ is saturated by 0.5 in (a) and 0.07 in (b) (even for $2m_S>m_h$) is
that the decay mode $h\rightarrow \phi \phi$ dominates
over the decay mode $h\rightarrow SS$ in this case (See Eq. (21)).
The lower curve indicates that we can rather easily probe the 2DM scenario by measuring the ratio $R$
as long as $2m_{\phi}<m_h$.

\section{Conclusion}

We have examined how the scenario of coexisting two-particle dark mater can be realized
in the extended seesaw model, which has been previously proposed
to accommodate small neutrino masses and low scale leptogenesis.
In this scenario, we impose the discrete symmetry $Z_2 \times Z_2^{\prime}$
and introduce new renormalizable interaction terms containing the heavy singlet scalar particle $\Phi$
so as for singlet Majorana neutrino $S$ and singlet scalar $\phi$ to be the coexisting two dark matter candidates.
Depending on the mass spectrum of the two dark matter candidates, the annihilation processes either
$SS\longrightarrow \phi \phi$ or $\phi\phi \longrightarrow SS$ is of particular interest because
the annihilation cross sections for the processes can be so large that the relic abundance of
decaying particle should get lowered, which in turn makes the constraints on its parameter space relaxed,
compared with the case of one and only one dark matter candidate.
We have also discussed the implications of the dark matter detection through the scattering
off the nucleus of the detecting material on our scenarios for dark matter candidate.
Our results show that the expected reach of SuperCDMS could detect a signal for dark matter or
at least strongly constrain the parameter space,
while the recent result of XENON10 Dark Matter experiment does not so.
In addition, we have studied the implications for the search
of invisible Higgs decay at LHC which may serve as a probe of our scenarios for dark matter.
In particular, we have found that there is a chance to easily probe the 2DM scenario by measuring
the ratio $R$ as long as $2 m_{\phi} <  m_h$.
\\

\noindent {\bf Acknowledgement:}
SKK is supported  by the KRF Grant funded by the
Korean Government(MOEHRD) (KRF-2006-003-C00069).
CSK and HSC are supported in part by CHEP-SRC and in part
by the Korea Research Foundation Grant funded by the Korean Government (MOEHRD)
No. KRF-2005-070-C00030.
\\

\end{document}